\documentstyle[aps,multicol,epsf,psfig]{revtex}

\newcommand{\beq}{\begin{equation}}
\newcommand{\eeq}{\end{equation}}
\newcommand{\beqa}{\begin{eqnarray}}
\newcommand{\eeqa}{\end{eqnarray}}
\newcommand{\ket} [1] {\vert #1 \rangle}
\newcommand{\bra} [1] {\langle #1 \vert} 

\begin{document}
\title{Optimal $N$-to-$M$ Cloning of Quantum Coherent States}
\author{N. J. Cerf$^{1,2}$ and S. Iblisdir$^1$}
\address{$^1$ Ecole Polytechnique, CP 165, Universit\'e Libre de Bruxelles,
B-1050 Bruxelles, Belgium\\
$^2$ Information and Computing Technologies Research Section,
Jet Propulsion Laboratory,\\
California Institute of Technology, Pasadena, CA 91109}

\date{May 2000}
\draft

\maketitle

\begin{abstract}
The cloning of continuous quantum variables is analyzed 
based on the concept of Gaussian cloning machines, i.e., transformations
that yield copies that are Gaussian mixtures centered on
the state to be copied. The optimality of Gaussian cloning machines 
that transform $N$ identical input states into $M$ output states 
is investigated, and bounds on the fidelity of the process are derived 
via a connection with quantum estimation theory.
In particular, the optimal $N$-to-$M$ cloning fidelity for coherent states 
is found to be equal to $MN/(MN+M-N)$.
\end{abstract}


\begin{multicols}{2}

Cloning denotes an operation by which the unknown state of a system is copied. 
For reasons rooted at the linearity of quantum mechanics, it turns out
that when the system to be copied is quantum mechanical, cloning cannot
be performed exactly\cite{bib_wzd}. Then, a natural question that arises
is ``to what extent can the copies resemble the original, in accordance with
quantum mechanics?''\cite{bib_bh}. 
This is the problem of optimal quantum cloning, 
which has now been extensively studied for quantum 
bits\cite{bib_gisinmassar,bib_brussetal,bib_gisin,bib_niu,bib_cerfprl},
and, more generally, for $d$-level 
systems\cite{bib_werner,bib_bhprl,bib_cerf}. 
The present paper investigates the question of optimal cloning 
for {\em continuous} quantum variables. 
Examples of continuous variables include
the position and momentum of a particle, or the two quadratures 
of a quantized electromagnetic field.
We will consider a quantum system described in terms of two canonically 
conjugate operators with continuous spectra (referred to as 
$\hat{x}$ and $\hat{p}$, with eigenvalues $x$ and $p$, respectively).
Precisely because they are conjugate, 
$\hat{x}$ and $\hat{p}$ cannot be both copied exactly. 
Nevertheless, approximate
cloning can be achieved if the copies are not required to be exact.
Then, the issue of optimal cloning amounts to
find the best tradeoff between position and momentum errors
induced by cloning.
\par

In this paper, we shall consider $N \to M$ symmetric Gaussian 
cloners (SGC), defined as a linear completely positive map 
$C_{N,M}$ transforming $N$ identical replicas of an unknown quantum state 
$\ket{\psi}$ belonging to an infinite-dimensional Hilbert space ${\cal H}$
into $M \geq N$ imperfect clones. The joint state of these clones 
$\rho_M=C_{N,M}(\ket{\psi^{\otimes N}}\bra{\psi^{\otimes N}})$ 
is required to be supported on 
the symmetric subspace of ${\cal H}^{\otimes M}$,
and is such that the partial trace over all outputs except one 
is the bi-variate Gaussian mixture
\beqa \label{eq_reducrho}
\rho_1 &=& {\mathrm Tr}_{M-1} (\rho_M) \nonumber \\
&=&\frac{1}{\pi \sigma^2_{N,M}}
\int d^2\beta \; e^{-|\beta|^2/\sigma^2_{N,M}} \;
D(\beta)\ket{\psi}\bra{\psi}D^{\dagger}(\beta)
\eeqa
where the integral is performed over all values of 
$\beta=(x+ip)/\sqrt{2}$ in the complex plane ($\hbar=1$), 
and the operator $D(\beta)=\exp(\beta {\hat a}^\dagger - \beta^* {\hat a})$ 
achieves a displacement of $x$ in position and $p$ in momentum,
with ${\hat a}$ and ${\hat a}^\dagger$ denoting 
the destruction and creation operators, 
respectively\cite{bib_wm}. Thus, the copies yielded by a SGC are affected
by an equal Gaussian noise $\sigma_x^2=\sigma_p^2=\sigma_{N,M}^2$ 
on the conjugate variables $x$ and $p$. (It will turn out that
the resulting cloning fidelity $f=\bra{\psi}\rho_1\ket{\psi}$
is invariant for all coherent states of $\hat{x}$ and $\hat{p}$.)
The symmetry of the cloner also obviously implies that the $M$ copies are
characterized each by the same density operator $\rho_1$.
\par

The issue of the duplication ($N=1$, $M=2$) of quantum information carried
by a continuous variable has been treated in a previous paper\cite{bib_cclon}, 
where an explicit Gaussian $1\to 2$ cloning transformation was proposed.
It was shown that the noise variance induced by this cloner is
$\sigma^2_{1,2}=1/2$, so that the resulting cloning fidelity
for coherent states is $f_{1,2}=2/3$. 
This fidelity is invariant under translations
and rotations in phase space, so that this Gaussian cloner can be thought of
as the analogue for coherent states of the universal cloning machine 
for quantum bits\cite{bib_bh}.
The present work investigates the optimality of this $1\to 2$ cloner, and
extends these considerations to $N\to M$ continuous cloners. More 
specifically, we address the question of ``how close'' the output
state [Eq. (\ref{eq_reducrho})] can be from 
the input state $\ket{\psi}$. We find that a
lower bound on the noise variance $\sigma_{N,M}^2$ is given by
\beq\label{eq_mainresult}
\overline{\sigma}^2_{N,M} =
\frac{M-N}{MN}
\eeq
implying in turn that the {\em optimal} $N\to M$ cloning fidelity 
for coherent states is bounded by
\beq  \label{eq_fidelity}
f_{N,M}={MN\over MN+M-N}
\eeq
\par

First, let us demonstrate that the bound (\ref{eq_mainresult})
is achieved with the $1 \to 2$ SGC derived in \cite{bib_cclon},
so that the latter is optimal for coherent states (optimality was only
conjectured in \cite{bib_cclon}). Our proof is directly connected
to the problem of simultaneously measuring a pair of conjugate observables
on a single quantum system. It is known (see e.g. \cite{bib_AK})
that any attempt to measure $\hat{x}$ and $\hat{p}$ simultaneously on a quantum
system is constrained by the inequality
\beq \label{eq_ak}
\sigma^2_x(1) \; \sigma^2_p(1) \geq 1
\eeq
where $\sigma^2_ x(N)$ and $\sigma^2_p(N) $ denote the variance of the 
measured values of $\hat{x}$ and $\hat{p}$, respectively, when $N$ replicas
of the state are available. (The case where $N>1$ will
be considered later on.) So, the best possible simultaneous
measurement of $\hat{x}$ and $\hat{p}$ with a same precision
satisfies $\sigma^2_x(1)=\sigma^2_p(1)=1$.
Compared with the intrinsic noise of a minimum-uncertainty wave packet 
$\sigma^2_x=\sigma^2_p=1/2$, we see that the joint measurement
of $x$ and $p$ effects an additional noise of minimum variance 
1/2 \cite{bib_AK}.
Now, let a coherent state $\ket{\alpha}$ be processed 
by a $1 \to 2$ SGC, and let $\hat{x}$ be 
measured at one output of the cloner while $\hat{p}$ is measured
at the other output. As cloning should obey inequality (\ref{eq_ak}), 
we must have
\beq
\Delta \hat{x}^2 \; \Delta \hat{p}^2 \geq 1
\eeq 
where $\Delta {\hat{x}}^2$ ($\Delta {\hat{p}}^2$) refers to the usual 
variance of observable $\hat{x}$ ($\hat{p}$) measured on $\rho_1$. 
Using Eq.~(\ref{eq_reducrho}), it gives
\beq
(\delta\hat{x}^2+\sigma_{1,2}^2) 
(\delta\hat{p}^2+\sigma_{1,2}^2) \geq 1
\eeq
where $\delta\hat{x}^2$ ($\delta\hat{p}^2$) is the intrinsic variance
of $\hat{x}$ ($\hat{p}$) measured on the input state,
while $\sigma_{1,2}^2$ is the noise variance induced by the cloner.
Now, using the uncertainty principle 
$\delta\hat{x}^2 \delta\hat{p}^2 \geq 1/4$ and the identity
$a^2+b^2\geq 2\sqrt{a^2 b^2}$, we
conclude that the noise variance is constrained by
\beq
\sigma^2_{1,2} \geq \overline{\sigma}^2_{1,2} = 1/2
\eeq
implying that the cloner presented in \cite{bib_cclon} is optimal. 
\par

Let us now consider the general problem of optimal $N \to M$ Gaussian cloning. 
Our proof is connected to quantum state estimation theory
similarly to what was done for quantum bits in \cite{bib_bruss},
the key idea being that cloning should not be a way of circumventing
the noise limitation encountered in any measuring process.
More specifically, our bound relies
on the fact that cascading a $N \to M$  cloner with a $M \to L$ cloner results 
in a $N\to L$ cloner which cannot be better that the
{\em optimal} $N \to L$ cloner. We make use of the property
that cascading two SGCs results in a single SGC whose variance 
is simply the sum of the variances of the two component SGCs
(see Appendix). Hence, the variance $\overline{\sigma}^2_{N,L}$ of 
the {\em optimal} $N \to L$ SGC must satisfy
$\overline{\sigma}^2_{N,L} \leq \sigma^2_{N,M}+\sigma^2_{M,L}$. 
In particular, if the $M\to L$ cloner is itself optimal and $L\to\infty$,
\beq\label{eq_addvar}
\overline{\sigma}^2_{N,\infty} \leq \sigma^2_{N,M}+
\overline{\sigma}^2_{M,\infty}
\eeq
Since the limit of $C_{N,M}$ with $M \to \infty$ corresponds to a 
measurement\cite{bib_gisinmassar}, Eq.~(\ref{eq_addvar})
implies that cloning the $N$ replicas of a system before measuring
the $M$ resulting clones
does not provide a mean to enhance the accuracy of a direct measurement
of the $N$ replicas.
\par

Let us now estimate $\overline{\sigma}^2_{N,\infty}$, that is, the variance 
of an optimal joint measurement of $\hat{x}$ and $\hat{p}$
on $N$ replicas of a system.
From quantum estimation theory \cite{bib_holevo}, 
we know that the variance of the measured values of $\hat{x}$ and $\hat{p}$
on a single system, respectively $\sigma^2_x(1)$ and $\sigma^2_p(1)$,
are constrained by
\beq\label{eq_holevobound}
g_x \sigma^2_x(1)+g_p \sigma^2_p(1) \geq 
g_x \delta \hat{x}^2+ g_p \delta \hat{p}^2 
+ \sqrt{g_x g_p} 
\eeq
for all values of the constants $g_x, g_p >0$.
Note that, for each value of $g_x$ and $g_p$, a specific POVM 
based on a resolution of identity in terms of
squeezed states (whose squeezing parameter $r$ is a function of
$g_x$ and $g_p$) achieves this bound (see \cite{bib_holevo}). 
Moreover, when measurement is performed 
on $N$ independent and identical systems, 
the r.~h.~s. of (\ref{eq_holevobound}) is reduced by a factor $N^{-1}$, as in  
classical statistics\cite{bib_helstrom}. 
So, applying $N$ times the optimal single-system POVM
is the best joint measurement when $N$ replicas are available since 
it yields $\sigma_x^2(N)=N^{-1}\sigma_x^2(1)$ and  
$\sigma_p^2(N)=N^{-1}\sigma_p^2(1)$. Hence, using Eq.~(\ref{eq_holevobound})
for a coherent state ($\delta \hat{x}^2=\delta \hat{p}^2=1/2$)
and requiring $\sigma_x^2(N)=\sigma_p^2(N)$,
the tightest bound is obtained for $g_x=g_p$. It yields
$\overline{\sigma}^2_{N,\infty}=1/N$, which, combined with 
Eq.~(\ref{eq_addvar}), gives the minimum noise 
variance induced by cloning, Eq.~(\ref{eq_mainresult}).
\par

It is now easy to compute the fidelity of the optimal $N \to M$ SGC 
when a coherent state $\ket{\alpha}$ is copied. Using Eq.~(\ref{eq_reducrho})
and the identity $|\langle \alpha|\alpha'\rangle|^2=\exp(-|\alpha-\alpha'|^2)$,
we obtain
\beq\label{eq_bestfid}
f_{N,M}
=\bra{\alpha}\rho_1\ket{\alpha}
= \frac{1}{1+\overline{\sigma}^2_{N,M}}
\eeq
which results in Eq.~(\ref{eq_fidelity}). As expected, 
all coherent states are copied with a same fidelity. (Note, however,
that this property does not extend to all states of ${\mathcal H}$.)
Equations (\ref{eq_mainresult}) and (\ref{eq_fidelity}) are consistent with 
the known result for a $1\to 2$ continuous cloner, i.~e., 
$\overline{\sigma}^2_{1,2}=1/2$ and $f_{1,2}=2/3$ \cite{bib_cclon}.
In addition, they yield the obvious result $\overline{\sigma}^2_{N,N}=0$
and $f_{N,N}=1$, confirming that the optimal $N\to N$ cloning map
is just the identity. Furthermore, they 
fulfill the natural requirement that the cloning fidelity increases with 
the number of input replicas. For instance, considering a
$kN \to kM$ SGC with a positive integer $k$, 
we find that $\frac{\partial \overline{\sigma}^2_{N,M}}{\partial k} <0$ 
(and $\frac{\partial f}{\partial k} >0$). At the limit
$N \to \infty$, we have $f \to 1$, $\forall M$, that is,
classical copying is allowed. 
Finally, for $M \to \infty$, that is, for an optimal measurement,
we get $f \to N/(N+1)$. In particular, it 
implies that the best simultaneous measurement of ${\hat x}$ and ${\hat p}$
on a single system gives a fidelity 1/2, a well-known result. 
\par

It is worth noting that optimally cloning squeezed states requires
a variant of these SGCs, just as in \cite{bib_cclon}.
Let us consider for instance a family of quadrature squeezed states
with squeezing parameter $r$.
For such a family, 
the best symmetric cloner must have the form of Eq.~(\ref{eq_reducrho}),
but using the definition $\beta=(\frac{x}{\sigma}+i\sigma p)/\sqrt{2}$
with $\sigma=\exp(r)$.
These cloners naturally generalize the SGCs and gives the same
cloning fidelity, Eq.~(\ref{eq_fidelity}), for those squeezed states.
\par

In conclusion, we have 
established a link between optimality of $N\to M$ symmetric Gaussian cloners 
and the impossibility of simultaneously measuring two conjugate observables 
$\hat{x}$ and $\hat{p}$. This results in a lower bound on the noise
induced by cloning. The optimal 
cloning fidelity for coherent states was then derived, and was found
to be independent of which coherent state is to be copied.
The optimal cloning of squeezed states was also found to be
equivalent to that of coherent states, as expected since 
the former can always be obtained by applying a canonical transformation
on the latter.  It is unknown whether a cloner specifically devised for other
classes of states might yield a fidelity 
exceeding Eq.~(\ref{eq_fidelity}). However,
since minimum-uncertainty states are the closest to 
classical states, we conjecture that SGCs achieve
the {\em best} possible fidelity if we require the cloner to be covariant
under rotations and translations in the phase space.
Finally, even though the explicit transformation 
achieving the $1\to 2$ optimal SGC is known\cite{bib_cclon},
finding the $N \to M$  cloning
transformation that attains the maximum fidelity 
is still an open question.
\par

\medskip

{\it Appendix.}
We now prove that the variances of two cascaded cloners add.
Consider a $N\to M$ SGC, followed by a $M\to L$ SGC. 
Let $\rho$ be an arbitrary density operator supported
on $\mathcal{H}^{\otimes M}$. Since it is self-adjoint and compact,
$\rho$ has a denumerable spectrum: it can be expanded as
$\rho=\sum_{i=1}^{\infty} \lambda_i \ket{\xi_i} \bra{\xi_i}$ with 
$\bra{\xi_i} \xi_j \rangle=\delta_{ij}$, $\lambda_i \ge 0$ and 
$\sum_{i=1}^\infty 
\lambda_i=1$. Note that $\forall \epsilon >0$, $\exists d$ such that 
$|\sum_{i=1}^{d} \lambda_i-1|<\epsilon$.
Therefore, the output of the first cloner can be decomposed as
$\rho_M = \rho_d+\epsilon_d \; B_d$
where $\rho_d=\sum_{i=1}^d \lambda_i \ket{\xi_i} \bra{\xi_i}$ is
supported on a $d$-dimensional subspace of 
$\mathcal{H}^{\otimes M}$, $B_d$ is a bounded operator, and 
$\lim_{d \to \infty} \epsilon_d=0 $. Since $\rho_M$ belongs to the symmetric 
subspace of $\mathcal{H}^{\otimes M}$, so will $\rho_d$.
Hence, we know that we can write 
$\rho_d$ in the form of a pseudo-mixture of pure product states 
$\rho_d=\sum_{i=1}^d \alpha_i 
\ket{\phi_i^{\otimes M}} \bra{\phi_i^{\otimes M}}$
where the coefficients $\alpha_i$ are not necessarily positive
but satisfy $\sum_{i=1}^d \alpha_i=1$ 
(see \cite{bib_werner} or \cite{bib_bruss}). 
Thus, when cloning a state 
$\ket{\psi^{\otimes N}}$, we have 
\beq  \label{eq_decomposition}
C_{N,M}(\ket{\psi^{\otimes N}}\bra{\psi^{\otimes N}})=
\sum_{i=1}^d \alpha_i 
\ket{\phi_i^{\otimes M}} \bra{\phi_i^{\otimes M}} + \epsilon_d \; B_d
\eeq
Then,
since the cloning map $C_{N,M}$ is linear, cascading the two cloners yields
$C_{M,L}C_{N,M}(\ket{\psi^{\otimes N}}\bra{\psi^{\otimes N}})=
\sum_i \alpha_i C_{M,L}(\ket{\phi_i^{\otimes N}} \bra{\phi_i^{\otimes N}})
+\epsilon_d C_{M,L}(B_d)$.
As this expression is a density operator (thus bounded) and the first term of 
its r.h.s. is positive, 
$C_{ML}(B_d)$ must be bounded. Thus, the second term of the
r.~h.~s. of Eq.~(\ref{eq_decomposition}) 
becomes negligible when $d\to \infty$. Now,
using  Eq.(\ref{eq_reducrho}), we have 
\beqa
\lefteqn{
{\mathrm Tr}_{L-1} 
C_{M,L}C_{N,M}(\ket{\psi^{\otimes N}}\bra{\psi^{\otimes N}})=
} \hspace{1truecm} \nonumber \\
& & {1\over \pi^2 \sigma_{M,L}^2 \sigma_{N,M}^2 }
\int d^2\gamma \; d^2\beta \; 
{\rm e}^{-|\gamma|^2 / \sigma^2_{M,L} 
-|\beta|^2 / \sigma^2_{N,M}} \nonumber \\
& & \times D(\gamma+\beta) \ket{\psi}\bra{\psi} D^{\dagger}(\gamma+\beta)
+O(\eta_d)
\eeqa
with $\lim_{d \to \infty}\eta_d=0$. A little algebra then shows 
that this last expression is a Gaussian mixture 
centered on the original state
whose variance is $\sigma^2_{M,L}+\sigma^2_{N,M}$. 
\par
\medskip


We are grateful to Serge Massar for very helpful discussions, 
especially concerning quantum estimation theory.
S. I. acknowledges support from the Fondation Universitaire
Van Buuren at the Universit\'e Libre de Bruxelles.

\end{multicols}
\end{document}